\documentstyle[aps,prb,twocolumn,epsf,rotate,graphics]{revtex}

\begin{document}
\twocolumn[\hsize\textwidth\columnwidth\hsize\csname
@twocolumnfalse\endcsname

\title{
Anomalous Hall effect in ferromagnetic semiconductors}

\draft

\author{T. Jungwirth$^{1,2}$,
Qian Niu$^{1}$, and A. H. MacDonald$^{1}$ 
}
\address{
$^{1}$Department of Physics, The University of Texas, Austin, TX 78712}
\address{
$^{2}$Institute of Physics ASCR, Cukrovarnick\'a 10,  
162 53 Praha 6, Czech Republic
}
\date{\today}
\maketitle

\begin{abstract}
We present a theory of the anomalous Hall effect in ferromagnetic
(Mn,III)V semiconductors.
Our theory relates the anomalous Hall conductance of a homogeneous ferromagnet 
to the Berry phase acquired by a quasiparticle wavefunction upon traversing
closed paths on the spin-split Fermi surface of a ferromagnetic state.
It can be applied equally well to any itinerant electron ferromagnet.
The quantitative agreement between our theory 
and experimental data in both (In,Mn)As and (Ga,Mn)As systems
suggests that this disorder independent contribution to the 
anomalous Hall conductivity dominates in diluted magnetic semiconductors.
\end{abstract}


\vskip2pc]

In recent years the semiconductor research community has enjoyed 
a remarkable achievement, making III-V compounds ferromagnetic by doping
them with magnetic elements.  The 1992 discovery\cite{ohnoprl92}
of hole-mediated ferromagnetic order in (In,Mn)As has 
motivated research on GaAs\cite{ohnoapl96} and other III-V host materials.
Ferromagnetic transition temperatures in excess of 100 Kelvin\cite{matsukuraprb98,ohnosci98}
and long spin-coherence times in GaAs\cite{kikkawanature99,malajovichnature01} 
have fueled hopes that a new magnetic medium is emerging that could open radically new  
pathways for information processing and storage technologies. The recent confirmation\cite{gamnntc}
of the room temperature ferromagnetism predicted\cite{dietlohno} in (Ga,Mn)N has added to
interest in this class of materials.  In both (In,Mn)As and (Ga,Mn)As systems, measurements of the anomalous Hall 
effect\cite{smitphysica58,luttingerpr58,bergerprb70,chienplenum80,chunprl00,crepieuxprb01} 
have played a key role in establishing ferromagnetism, 
and in providing evidence for the essential role of hole-mediated coupling between Mn 
local moments in establishing long-range order.\cite{ohnoprl92,ohnoapl96,ohnojmmm99} 
Despite the importance of the anomalous Hall effect (AHE) for sample
characterization, a theory which allows these experiments to be interpreted quantitatively 
has not been available.  In this article we present a theory of the AHE in 
ferromagnetic III-V semiconductors that appears to account for existing observations.

The Hall resistivity of ferromagnets has an ordinary contribution,
proportional to the external magnetic field strength, and an anomalous
contribution usually assumed to be proportional
to the sample magnetization.
The classical theory of the anomalous Hall effect (AHE) in a metal\cite{chienplenum80} 
starts from the mean-field Stoner theory description of its ferromagnetic state, in which 
current is carried by quasiparticles in spontaneously
spin-split Bloch bands.  A similar mean-field theory 
has recently been developed\cite{jungwirthprb99,dietlsci00,abolfathprb01,dietlprb01} 
and used to interpret magnetic 
properties of (III,Mn)V ferromagnets.  In these theories 
the host semiconductor valence bands are split by 
an effective field that results from exchange interactions
with polarized Mn  moments.  The field makes a
wavevector independent contribution,
\begin{equation}
H_{split} =  h \; \hat m \cdot \vec s
\label{hmt}
\end{equation}
to the band Hamiltonian.  Here $\hat m$ is the polarization direction of the 
local moments  
and $\vec s$ is the electron
spin-operator.  In the (In,Mn)As and (Ga,Mn)As AHE measurements,\cite{ohnoprl92,ohnoapl96}
$\hat m$ is in the $\langle 00\overline{1} \rangle$ direction for positive external
magnetic fields. The effective field $h$ is proportional
to the average local moment magnetization and is non-zero only in
the ferromagnetic state. The antiferromagnetic 
interaction\cite{dietlhand94,okabayashiprb98}
between localized and itinerant spins implies that $h>0$. 
When  
Mn spins are fully polarized, $h= N_{Mn}SJ_{pd}$, where
$N_{Mn}$ is the density of Mn ions with spin  
$S=5/2$ and $J_{pd}=50\pm5$~meVnm$^3$ is
the strength of the exchange coupling 
between the local moments and the valence band electrons.\cite{ohnojmmm99}
From a symmetry point of view, the AHE is 
made possible by this effective magnetic field, and by
the spin-orbit coupling present in the host semiconductor 
valence band.  

In the standard model of the AHE in metals, 
skew-scattering\cite{smitphysica58} and side-jump\cite{bergerprb70} scattering
give rise to contributions to the Hall resistivity 
proportional to the diagonal resistivity $\rho$ 
and $\rho^2$ respectively, with the latter process tending to 
dominate in alloys because $\rho$ is larger.
Our evaluation of the AHE in (III,Mn)V
ferromagnets is based on a theory of semiclassical
wave-packet dynamics, developed previously by one of 
us\cite{sundaramprb99} which 
implies a contribution to the Hall conductivity 
independent of the kinetic equation scattering term.  Our focus on this 
contribution is motivated in part by practical considerations, since our
current understanding of (III,Mn)V ferromagnets is not sufficient
to permit confident modeling of quasiparticle scattering.  
The relation of our approach to standard theory is reminiscent 
of disagreements between Smit\cite{smitphysica58} and Luttinger\cite{luttingerpr58} 
that arose early in the development of AHE theory and do not appear 
to have ever been fully resolved.  In this paper we follow
Luttinger\cite{luttingerpr58} in taking the view that there is a contribution
to the AHE due to the change in wavepacket group velocity that occurs
when an electric field is applied to a ferromagnet.  
Since the Hall resistivity is invariably smaller than the diagonal resistivity,
a temperature independent value of the Hall conductivity corresponds to a Hall 
resistivity proportional to $\rho^2$, usually interpreted as evidence for 
dominant side-jump scattering.  As we explained below, we find quantitative
agreement between our Hall conductance values and experiment, suggesting that 
the AHE {\em conductance} value may be intrinsic in many metallic ferromagnets.

The Bloch electron group velocity correction is conveniently evaluated using expressions derived
by Sundaram and Niu\cite{sundaramprb99}:
\begin{equation}
\dot x_{c} = \frac{\partial \epsilon}{\hbar \partial \vec k}
+ (e/\hbar) \vec E \times \vec \Omega.
\label{velocity}
\end{equation}
The first term on the right-hand-side of Eq.~\ref{velocity} is 
the standard Bloch band group velocity.  Our anomalous Hall conductivity 
is due to the second term, proportional to the $\vec k$-space Berry curvature
$\vec{\Omega}$.  
It follows from symmetry considerations that for a cubic semiconductor 
under lattice-matching strains and with $\hat m$ aligned by
external fields along the $\langle 001\rangle$ growth direction, only $\Omega_{z} \ne 0$:
\begin{equation}
\Omega_z(n,\vec k) = 2{\rm Im}
 \big[ \langle \frac{\partial u_n}{\partial k_y} \vert \frac{\partial 
u_n}{\partial k_x} \rangle 
\big].
\label{omega}
\end{equation}
Here $|u_n\rangle$ is the periodic part of the $n$-th Bloch band
wavefunction with the mean-field spin-splitting term included in the Hamiltonian.
The anomalous Hall conductivity that results from this velocity correction is
\begin{equation}
\sigma_{AH} = -\frac{e^2}{\hbar} \sum_{n} \int \frac{d \vec k}{(2 \pi)^3} 
 f_{n,\vec k} \Omega_{z}(n,\vec k)\; ,
\label{sigmaH}
\end{equation}
where 
$f_{n,\vec k}$ is the equilibrium Fermi occupation factor for the band
quasiparticles.  We have taken the convention that a positive $\sigma_{AH}$
means that the anomalous Hall current is in the same direction as the normal Hall current.  

This Berry phase contribution occurs for any itinerant electron ferromagnet. 
To assess its importance for (III,Mn)V compounds, we first explore a simplified model that
yields parabolic dispersion
for both heavy--hole and light--hole bands and a spin-orbit coupling\cite{laserbook,abolfathprb01} 
strength that is much larger than the hole Fermi energy.
Detailed numerical calculations  
that account for mixing of the spin-orbit split-off bands and warping of the occupied
heavy--hole and light--hole bands\cite{laserbook,abolfathprb01} 
in (In.Mn)As and (Ga,Mn)As samples\cite{ohnoprl92,matsukuraprb98}
will follow this general and qualitative discussion.  Within a 4-band model, the spin operator
$\vec s=\vec j/3$ in Eq.(\ref{hmt}), and the spherical model Hamiltonian for holes in III-V host semiconductors
can be written as

\begin{equation}
H_0=\frac{\hbar^2}{2m}
\big[(\gamma_1+\frac52\gamma_2)k^2-2\gamma_2(\vec k\cdot\vec j)^2\big]\, 
{\rm Ry}a_0^2
\; ,
\label{h0}
\end{equation}
where 
$\vec j$ is the total angular momentum operator, 
$\gamma_1$ and $\gamma_2$ are Luttinger parameters,\cite{laserbook,iiiv}
and $a_0$ is the Bohr radius.
In the unpolarized case ($h=0$), the total Hamiltonian, $H=H_0+H_{split}$, is diagonalized 
by spinors $|j_{\hat k}\rangle$ where, e.g.,  $j_{\hat k}\equiv \vec j\cdot\hat k=\pm3/2$ for 
the two degenerate heavy--hole
bands with effective mass $m_{hh}=m/(\gamma_1-2\gamma_2)$.      
The Berry phase is familiar in this case since the Bloch eigenstates are $j=3/2$ spin
coherent states\cite{Auerbach}.   Integrating over planes of occupied states at fixed $k_z$ we find that 
$\int d^2k f_{n,\vec k} \Omega_{z}(n,\vec k) =\pm3/2(\cos\theta_{\vec k} -1)$ 
where $\cos(\theta_{\vec k}) 
\equiv k_z/k_{hh}$ and $k_{hh}$ is the Fermi wavevector.
The anomalous Hall conductivity (\ref{sigmaH}) vanishes in the $h=0$ limit because 
the contributions from the two heavy hole bands, and also from the two light hole bands,
cancel.  In the ferromagnetic state, on the other hand,
majority and minority spin heavy and light hole Fermi surfaces 
differ and also  the Berry phases are modified when $h \ne 0$.
Up to linear order in $h$
we obtain that $k_{hh}^{\pm}=k_{hh}\pm \cos\theta_{\vec k} 
hm_{hh}/(2\hbar^2k_{hh})$
and  the Berry phase is altered by
the factor $(1\mp 2mh/(9\gamma_2\hbar^2k_{hh}^2)$.  
A similar analysis for the light-hole bands leads to
a total net contribution to the AHE from the four bands whose
lower and upper bounds are:

\begin{eqnarray}
& &\frac{e^2}{2\pi\hbar}\frac{h}{2\pi\hbar^2}(3\pi^2p)^{-1/3}m_{hh}<\sigma_{AH}<
\nonumber \\
& &\frac{e^2}{2\pi\hbar}\frac{h}{2\pi\hbar^2}(3\pi^2p)^{-1/3}2^{2/3}m_{hh}\;.
\label{sigmaH4b}
\end{eqnarray}
Here $p=k_{hh}^3/3\pi^2\, (1+\sqrt{m_{lh}/m_{hh}})$ is the total hole density
and $m_{lh}=m/(\gamma_1+2\gamma_2)$ is the light-hole
effective mass. The lower bound in Eq.(\ref{sigmaH4b}) is obtained assuming 
$m_{lh}<<m_{hh}$ while the upper bound is reached when $m_{lh}\approx m_{hh}$.

\begin{figure}
\epsfxsize=3.3in
\hspace*{-0.5cm}
\centerline{
\epsffile{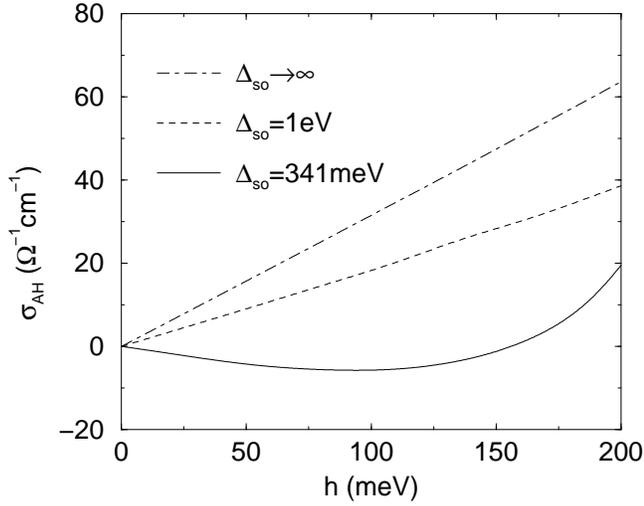}}

\vspace*{.7cm}   

\caption{  
Illustrative calculations of the 
anomalous Hall conductance as a function of polarized Mn ions field for hole
density $p=0.35$~nm$^{-1}$. The dotted-dashed curve was obtained
assuming infinitely large spin-orbit
coupling and the decrease of theoretical $\sigma_{AH}$ with decreasing
spin-orbit
coupling strength is demonstrated for $\Delta_{so}=1$~eV (dashed line) and $\Delta_{so}=341$~meV
(solid line).
}
\label{fig1}
\end{figure}

Based on the above analysis we conclude that the Berry phase anomalous
velocity can yield a sizeable AHE in (III,Mn)V ferromagnets.
The solid line in Fig.\ref{fig1} shows our analytic results for
GaAs effective masses $m_{hh}=0.5m$ and $m_{lh}=0.08m$. Note that in experiment,
anomalous Hall conductances are in order of 1--10 ${\rm \Omega^{-1}\, cm^{-1}}$  
and the  effective exchange field $h\sim 10-100$~meV.  According to Eq.~(\ref{sigmaH4b})
larger $\sigma_{AH}$ values should be expected in systems with larger heavy--hole effective
masses and in systems with the ratio $m_{lh}/m_{hh}$ close to unity. 

\begin{figure}
\epsfxsize=3.3in
\hspace*{-0.5cm}
\centerline{
\epsffile{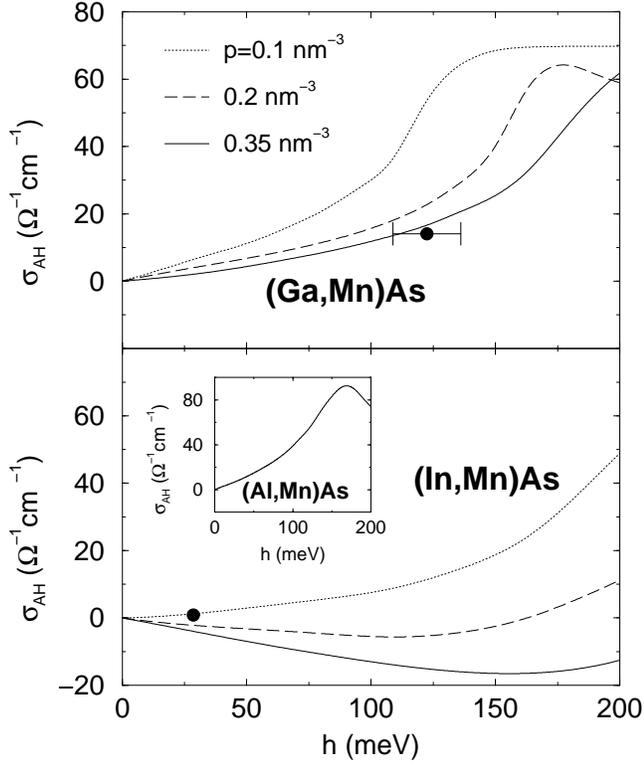}}  

\vspace*{.7cm}   

\caption{  
Full numerical simulations of $\sigma_{AH}$ for GaAs host (top panel), InAs host (bottom panel), and
AlAs host (inset) with hole densities $p=0.1$~nm$^{-1}$ (dotted lines), $p=0.2$~nm$^{-1}$ (dashed lines),
and $p=0.35$~nm$^{-1}$ (solid lines). Following Luttinger parameters of the valence bands were 
used:\protect\cite{iiiv}
GaAs -- $\gamma_1=6.98$, $\gamma_2=2.06$, $\gamma_3=2.93$, $\Delta_{so}=341$~meV;
InAs -- $\gamma_1=20$, $\gamma_2=8.5$, $\gamma_3=9.2$, $\Delta_{so}=390$~meV;
AlAs -- $\gamma_1=3.76$, $\gamma_2=0.82$, $\gamma_3=1.42$, $\Delta_{so}=280$~meV.
Filled circles in the top and bottom panels represent mesured 
AHE.\protect\cite{ohnoprl92,matsukuraprb98,ohnojmmm99} The saturation
mean-field $h$ values for the two points
were estimated from nominal sample 
parameters.\protect\cite{ohnoprl92,matsukuraprb98,ohnojmmm99} Horizontal error bars correspond to the
experimental uncertainty of the $J_{pd}$ coupling constant. Experimental hole density
in the (Ga,Mn)As sample is $p=0.35$~nm$^{-1}$; for (In,Mn)As, $p=0.1$~nm$^{-1}$ was determined
indirectly from sample's transition temperature. 
}
\label{fig2}
\end{figure}

So far we have discussed
the limit of infinitely strong spin-orbit coupling with an exchange field that is 
small relative to the hole Fermi energy. In the opposite
limits of zero spin-orbit coupling or large $h$, $\sigma_{AH}$ vanishes. 
This implies that the anomalous Hall conductivity is generally nonlinear in the 
exchange field and the magnetization.  To explore the intermediate regime
we numerically diagonalized 
the 6-band Luttinger Hamiltonian\cite{laserbook,abolfathprb01} with the spin-orbit gap
$\Delta_{so}=1$~eV as well as for the GaAs value $\Delta_{so}=341$~meV.  The results
shown in Fig.\ref{fig1} confirm that smaller $\sigma_{AH}$ is expected
in systems with smaller $\Delta_{so}$ and suggest that both positive and negative signs
of $\sigma_{AH}$ can occur in general. The curves in Fig.\ref{fig1} 
are obtained by neglecting
band warping in III-V semiconductor compounds. The property that the valence bands in these materials
are strongly non-parabolic, even in the absence of the field $h$ and even in
the large $\Delta_{so}$ limit,
is accurately captured by introducing the third phenomenological Luttinger parameter
$\gamma_3$.\cite{laserbook,abolfathprb01} Our numerical results 
indicate that warping leads to an increase of $\sigma_{AH}$, as seen when comparing
the solid curves in Fig.\ref{fig1} and in the top panel of Fig.\ref{fig2}.
The hole-density dependence of $\sigma_{AH}$, illustrated in Fig.\ref{fig2},
is qualitatively consistent with the spherical model prediction (\ref{sigmaH4b}).
Also, in accord with the chemical trends outlined above, the numerical data in Fig.\ref{fig2}
suggest large positive AHE coefficient for (Al,Mn)As, an intermediate positive $\sigma_{AH}$ in (Ga,Mn)As,
and a relatively weak AHE in (In,Mn)As with a sign that may be sensitive to strain and other details
of a particular sample. 

We now compare our $\sigma_{AH}$ theory with the experimental data available in 
(In,Mn)As and (Ga,Mn)As samples, studied extensively by Ohno and
coworkers.\cite{ohnoprl92,matsukuraprb98,ohnosci98,ohnojmmm99} The
nominal Mn densities in these two systems are $N_{Mn}=0.23$~nm$^{-3}$ for
the InAs host and $N_{Mn}=1.1$~nm$^{-3}$ for the GaAs host, 
yielding saturation values of the effective field $h=25\pm3$~meV and $h=122\pm14$~meV,
respectively.  The low-temperature hole density of the (Ga,Mn)As sample,
$p=0.35$~nm$^{-3}$, was unambiguously determined \cite{ohnojmmm99}
from the ordinary Hall coefficient measured at high magnetic fields.
Since similar experiments have not been reported for the (In,Mn)As
sample, we estimated the hole density, $p=0.1$~nm$^{-3}$, by fitting the density-dependent 
mean-field theory $T_c$ to the measured value $T_c=7.5$~K.
The use of a mean-field theory description of the ferromagnetic 
state in both samples is justified\cite{konigprl00,schliemannapl01} by the homogeneity of the  samples
and by the relatively small Fermi energy density of states.
Indeed, the measured ferromagnetic transition temperature
for the (Ga,Mn)As sample, $T_c=110$~K, is in an excellent agreement with the calculated
transition temperature\cite{dietlsci00,jungwirthphysicae01}, and 
mean-field theory also successfully explains the magnetic anisotropy of  
both systems.\cite{dietlsci00,abolfathprb01} 
Luttinger parameters for the two host semiconductors are well known\cite{iiiv} 
and are listed in the caption of Fig.\ref{fig2}.
As demonstrated in Fig.\ref{fig2}, our theory explains the order of
magnitude difference between AHE's in the two materials ($\sigma_{AH}
\approx 1\,\Omega^{-1}$~cm$^{-1}$ in (In,Mn)As and 
$\sigma_{AH} \approx 14\,\Omega^{-1}$~cm$^{-1}$ in (Ga,Mn)As). The calculations are also 
consistent with the positive sign and monotonic dependences of $\sigma_{AH}$
on sample magnetizations.\cite{ohnojmmm99}

We take the  agreement in both magnitude and sign of the AHE 
as a strong indication that the anomalous
velocity contribution dominates AHE in homogeneous (III,Mn)V ferromagnets.
This Berry phase conductivity, which is independent of quasiparticle
scatterers, is relatively easily evaluated with high accuracy.
According to our theory, comparison of theoretical and experimental Hall conductivity values 
provides information not only on the magnetization but also on the character of the 
itinerant electron wavefunctions that participate in the magnetism.  For example,
we predict that size quantization effects in quantum wells that inhibit heavy-light hole mixing
will reduce the $\vec k$-space Berry curvatures and hence anomalous Hall conductivities.
The success reported here motivates a reexamination of this effect 
in all itinerant electron ferromagnets.

\section*{Acknowledgments}
We are grateful for helpful discussions with T. Dietl, J. Furdyna, F. Matsukura,
and H. Ohno.  Our work was supported by DARPA, the Indiana 21st Century Fund, the Welch 
Foundation, the Ministry of Education
of the Czech Republic, and the Grant Agency of the Czech Republic.

\begin{references}

\bibitem{ohnoprl92}
H. Ohno {\em et al.},
Phys. Rev. Lett. {\bf 68},
2664 (1992).

\bibitem{ohnoapl96}
H. Ohno {\em et al.},
Appl. Phys. Lett. {\bf 69}, 363 (1996).

\bibitem{matsukuraprb98} F. Matsukura {\em et al.},
Phys. Rev. B {\bf 57}, R2037 (1998).

\bibitem{ohnosci98}
H. Ohno, Science
{\bf 281}, 951 (1998).

\bibitem{kikkawanature99} J.M. Kikkawa and D.D. Awschalom, 
Nature {\bf 397}, 139 (1999).

\bibitem{malajovichnature01}
I. Malajovich {\em et al.}, 
Nature {\bf 411}, 770 (2001).

\bibitem{gamnntc} S. Sonoda {\em et al.},
e-print, [http://arXiv.org/abs/cond-mat/0108159]

\bibitem{dietlohno} T. Diet {\em et al.},
Science {\bf 287}, 1019 (2000).

\bibitem{smitphysica58} J. Smit,
Physica {\bf 23}, 39 (1958).


\bibitem{luttingerpr58} J.M. Luttinger,
Phys. Rev. {\bf 112}, 739 (1958).

\bibitem{bergerprb70}
L. Berger, 
Phys. Rev. B {\bf 2}, 4559 (1970).

\bibitem{chienplenum80}   
{\em The Hall Effect and Its Applications} (eds Chien, C.L. \& 
Westgate, C.R.) (Plenum, New York, 1980).

\bibitem{chunprl00} S.H. Chun {\em et al.},
Phys. Rev. Lett.  {\bf 84}, 757 (2000). 

\bibitem{crepieuxprb01}
A. Crépieux and P. Bruno,  
Phys. Rev. B
{\bf 64}, 014416-1 (2001).

\bibitem{ohnojmmm99}
H. Ohno, 
J. Magn. Magn. Mater. {\bf 200}, 110 (1999).

\bibitem{jungwirthprb99}  
T. Jungwirth {\em et al.},
Phys. Rev.
B {\bf  59},  9818  (1999).

\bibitem{dietlsci00} 
T. Dietl {\em et al.},
Science {\bf 287}, 1019 (2000). 

\bibitem{abolfathprb01}
M. Abolfath {\em et al.},
Phys. Rev. B {\bf 63}, 054418-1 (2001).

\bibitem{dietlprb01}
T. Dietl, H. Ohno, and F. Matsukura,
Phys. Rev. B {\bf 63}, 195205-1 (2001).

\bibitem{dietlhand94} 
T. Dietl in {\em Handbook of
Semiconductors} {\bf 3B}, 1264--1267 (North-Holland, Amsterdam, 1994).


\bibitem{okabayashiprb98} J. Okabayashi {\em et al.},
Phys. Rev. B {\bf 58}, R4211 (1998).

\bibitem{sundaramprb99} G. Sundaram and Q. Niu,
Phys. Rev. B {\bf 59},
14915 (1999). 

\bibitem{laserbook}
W.W. Chow, S.W. Koch, and M. Sargent III,
{\em Semiconductor laser physics}, 179--192 (Springer-Verlag, Berlin, 1999).


\bibitem{iiiv} I. Vurgaftman,  
J.R. Meyer, and L.R. Ram-Mohan, 
Appl. Phys. Rev., in press.

\bibitem{Auerbach} A. Auerbach {\it Interacting Electrons and Quantum Magnetism},
(Springer-Verlag, New York, 1994).

\bibitem{konigprl00}
J. K\"{o}nig, H.H. Lin, and A.H. MacDonald,
Phys. Rev. Lett. {\bf 84}, 5628 (2000).

\bibitem{schliemannapl01} J. Schliemann {\em et al.},
Appl. Phys. Lett. {\bf 78}, 1550 (2001).   

\bibitem{jungwirthphysicae01} T. Jungwirth and A.H. MacDonald,
Physica E {\bf 10}, 153 (2001).
\end{references}
\end{document}